\documentclass[a4paper,11pt]{article}
\usepackage{graphicx,amssymb,bm,latexsym}
\makeatletter
\@addtoreset{equation}{section}

\makeatother

\newcommand{\bea}{\begin{eqnarray}}
\newcommand{\ena}{\end{eqnarray}}
\newcommand{\vs}[1]{\vspace{#1 mm}}
\newcommand{\hs}[1]{\hspace{#1 mm}}
\renewcommand{\a}{\alpha}
\renewcommand{\b}{\beta}
\renewcommand{\c}{\gamma}
\newcommand{\G}{\Gamma}
\renewcommand{\d}{\delta}
\newcommand{\e}{\epsilon}
\newcommand{\s}{\sigma}

\newcommand{\la}{\lambda}
\newcommand{\pa}{\partial}

\newcommand{\nn}{\nonumber\\}
\newcommand{\p}[1]{(\ref{#1})}

\begin{document}

\begin{titlepage}

\begin{flushright}
KU-TP 052 \\
\end{flushright}

\vs{10}
\begin{center}
{\Large\bf A Complete Classification of Higher Derivative Gravity in $3D$ \\
and Criticality in $4D$}
\vs{15}

{\large
Nobuyoshi Ohta\footnote{e-mail address: ohtan@phys.kindai.ac.jp}} \\
\vs{10}
{\em Department of Physics, Kinki University,
Higashi-Osaka, Osaka 577-8502, Japan}

\vs{15}
{\bf Abstract}
\end{center}

We study the condition that the theory is unitary and stable in three-dimensional
gravity with most general quadratic curvature, Lorentz-Chern-Simons and cosmological terms.
We provide the complete classification of the unitary theories around flat Minkowski
and (anti-)de Sitter spacetimes.
The analysis is performed by examining the quadratic fluctuations around these
classical vacua.
We also discuss how to understand critical condition for four-dimensional theories
at the Lagrangian level.

\end{titlepage}
\newpage
\setcounter{page}{2}

\section{Introduction}

The quest for a quantum theory of gravity is one of the long standing problems
in theoretical physics. The usual Einstein gravity suffers from the problem
that the theory is non-renormalizable in four and higher dimensions. The addition
of higher derivative terms such as Ricci and scalar curvature squared terms makes
the theory renormalizable at the cost of the loss of unitarity~\cite{Stelle1}.
Of course, unitarity is quite important in any physical theory.
Otherwise the theory does not make sense.

Recently a very interesting proposal has been made that the addition of such higher
order terms to three-dimensional gravity makes the theory unitary and possibly renormalizable
if the coefficients are chosen appropriately~\cite{BHT1}. The usual Einstein gravity
does not have any propagating mode, but the addition of these terms introduces
propagating massive graviton around flat Minkowski and curved maximally symmetric
spacetimes [anti-de Sitter (AdS) and de Sitter (dS) spacetimes].
The theory of massive graviton with Lorentz-Chern-Simons (LCS) term has
long been known as topologically massive theory~\cite{Top}, but the theory violates parity.
In contrast, the new theory is a parity preserving theory, and called new massive gravity.
This is very interesting in that we have really dynamical theory of gravity that is
unitary even though higher derivative terms are included. Since then, various aspects
of the theory have been investigated.
Linearized excitations in the field equations were studied in~\cite{LS}.
Unitarity and renormalizability is studied in~\cite{Oda}, though the issue of
renormalizability is not on the firm foundation~\cite{Deser,BHTE}.
Unitarity is proven for Minkowski spacetime in~\cite{Deser,GST}, whereas
it is discussed in \cite{BHT2} for maximally symmetric spacetimes.
Supergravity extension is discussed in~\cite{BHT3,BHT4}.
The critical case is studied in~\cite{BHT5}.
The partial result of unitarity condition on the flat Minkowski spacetime was known
for the usual sign of the Einstein theory~\cite{NR}.
Related discussion based on the AdS/CFT correspondence is given in \cite{S1}.

Though this kind of theories have their own significance, it is also known that
such higher order terms are present in the low-energy effective theories of
superstrings. There is some ambiguity in such theories due to the field redefinition.
If the above approach of requiring unitarity determines the coefficients to certain
extent, it may cast some light to the superstrings themselves.

The theory in question contains Einstein, Ricci tensor squared, scalar curvature squared
terms as well as LCS term. The analysis is made for what range
of these parameters the theory is unitary on flat Minkowski spacetime
in \cite{BHT3,Deser,GST}, and on maximally symmetric spacetimes in \cite{BHT2}.
(For dS case, see also \cite{GST}.) To understand the problem of unitarity,
the analysis of field equations is not enough.
We should look at the quadratic fluctuations of the theory around possible vacua
in the Lagrangian and check if the physical particles have correct sign for kinetic terms.
This off-shell analysis has been made in \cite{Deser} for the theory around Minkowski vacuum,
in \cite{GST,BHT3} for that with LCS term around Minkowski, and
in \cite{BHT2} for that around maximally symmetric spacetimes but without LCS term
and with a particular relation between some coefficients obtained for Minkowski
spacetime from the outset. However whether the theory makes sense or not should be
studied for each vacuum; if it is unitary around a vacuum but not on the other,
we should simply consider the theory near the sensible vacuum.
Thus the complete classification of the unitary and stable theory (for which range
of parameters the theory is unitary and stable) including LCS term with arbitrary
coefficients for the theory on the maximally symmetric spacetimes is lacking.
One of the purposes of the present paper is to fill this gap and provide a complete
classification of possible unitary and stable theory for three-dimensional gravity,
with all terms and arbitrary coefficients, by looking at the quadratic fluctuation of
the theory around these vacua in the Lagrangian.
In this way, we also resolve a problem left unresolved in Ref.~\cite{BHT5} if the unitarity of
the AdS irreps is enough to ensure the unitarity of the field theory;
the answer is negative, and we are able to identify which range of the parameters
allows unitary and stable theory. This can be done only in the off-shell analysis.
We also find that there is certain parameter region that has not been explored before.

When the theory is considered around maximally symmetric spacetimes (including Minkowski
spacetime), it turns out that the theory is unitary and stable for the ``wrong'' sign of
the Einstein term if the Ricci tensor squared is present.
This means that the Einstein gravity may not be obtained in the low-energy
approximation to the theory in that case.
The question then arises if this could be remedied, and also what happens in
the four-dimensional case.
In this connection, three-dimensional critical theory was proposed with
the usual sign for the Einstein term, LCS and cosmological constant terms~\cite{LSS},
and it was argued that the unitarity might be recovered in the critical case with
a particular relation between cosmological constant and the mass term.
Motivated by this, an interesting proposal of critical gravity in four and
higher dimensions has been made~\cite{LP,DLLP}. It has been suggested that
in this critical theory the theory may be unitary by imposing suitable boundary
conditions to eliminate some modes. However there arise additional logarithmic
modes~\cite{GJ,Log} and it has been pointed out that there may be a trouble with
the unitarity~\cite{PR}. Also even if ghost modes can be eliminated by boundary
conditions, it still remains a question whether the theory makes sense or not
at the quantum level. This issue has to be examined further.
Most of the study so far are based on the field equations,
but once again we emphasize that the off-shell approach is the most suitable way
to study this problem. In view of this situation, we attempt to understand
this problem by extending our analysis in three dimensions to four dimensions.
We are informed that a related discussion is given in the appendix of Ref.~\cite{SGT}.

We should mention that some extensions of critical gravity to include further
higher order terms~\cite{LPP} and supergravity~\cite{LPSW} have been considered.
See also~\cite{MM}.

This paper is organized as follows.
In sect.~2, we present the general theory we study in this paper with
Einstein, scalar curvature squared, Ricci squared, cosmological and topological
mass terms with arbitrary coefficients.
We then proceed to the study of the condition for the unitarity and stability
around flat Minkowski spacetime in sect.~3 and that around maximally symmetric spacetimes
in sect.~4 by examining the quadratic fluctuations in the Lagrangian.
We find the complete conditions for each case, and also examine the critical
conditions.
In sect.~4, we present some formulae for studying the quadratic fluctuations
in general dimensions, and use them to discuss critical conditions in four dimensions.
In this way we get some new view on the criticality in gravitational theory.
Sect.~5 is devoted to our conclusions and discussions.
An appendix collects some useful formulae necessary in the text.

\section{The Most General Theory}

We consider the action
\bea
S=\frac{1}{\kappa^2}\int d^3x \Bigg\{ \sqrt{- g} \Big[ \s R - 2 \Lambda_0 + \a R^2
+\b R_{\mu\nu}^2 \Big] + {\cal L}_{LCS} \Bigg\},
\label{action}
\ena
where $\kappa^2$ is the three-dimensional gravitational constant,
$\a, \b, \mu$ and $\s (=0, \pm 1)$ are constants, $\Lambda_0$ is a cosmological constant,
and the last term is the LCS term
\bea
{\cal L}_{LCS}
= \frac{1}{2\mu} \e^{\mu\nu\rho}( \G^\a_{\mu\b} \pa_\nu \G^\b_{\rho\a}
+\frac{2}{3}\G^\a_{\mu\c}\G^\c_{\nu\b}\G^\b_{\rho\a}),
\label{cs}
\ena
where $\G$ is the usual Levi-Civita connection for the spacetime metric $g$.
Our conventions are summarized in the appendix.

The variation of each term gives the field equations:
\bea
\s G_{\mu\nu} +\Lambda_0 g_{\mu\nu} + \a E_{\mu\nu}^{(1)} + \b E_{\mu\nu}^{(2)}
+ \frac{1}{\mu} C_{\mu\nu}=0,
\ena
where
\bea
G_{\mu\nu} &=& R_{\mu\nu}-\frac12 R g_{\mu\nu}, \nn
E_{\mu\nu}^{(1)} &=& 2RR_{\mu\nu}-2\nabla_\mu \nabla_\nu R
 + g_{\mu\nu}\Big(2\Box R -\frac12 R^2\Big), \nn
E_{\mu\nu}^{(2)} &=& 2R_{\mu\la}R_\nu^\la - 2 \nabla^\la \nabla_{(\mu} R_{\nu)\la}
 + \Box R_{\mu\nu} + \frac12 (\Box R -R_{\la\rho}^2 )g_{\mu\nu}, \nn
C_{\mu\nu} &=& \e_\mu{}^{\a\b} \nabla_\a \Big(R_{\b\nu}-\frac14 g_{\b\nu} R \Big).
\ena
$G_{\mu\nu}$ and $C_{\mu\nu}$ are known as Einstein and Cotton tensors, respectively.

There are two possible vacua in the theory: Minkowski and maximally symmetric spacetimes
of (anti)-de Sitter ((A)dS).
Here we wish to study the range of the coefficients for which the theory is unitary
(no ghost) and stable. There have been several studies, but as far as we are aware,
there is no study of the system with the most general parameters.

We consider the action up to second order around the background spacetime
\bea
g_{\mu\nu}= \bar g_{\mu\nu} + \kappa h_{\mu\nu},
\label{fluc}
\ena
where the background $\bar g_{\mu\nu}$ is chosen to be a maximally symmetric
spacetime with Riemann curvature
\bea
\bar R^\a{}_{\b\mu\nu}
= \Lambda (\bar g^\a_{\mu}\bar g_{\b\nu}-\bar g^\a_{\nu}\bar g_{\b\mu}),
\label{back}
\ena
with Minkowski spacetime corresponding to $\Lambda=0$.
We define
\bea
h \equiv \bar g^{\mu\nu} h_{\mu\nu}, \qquad
h_\mu \equiv \nabla^\nu h_{\nu\mu}.
\ena
Here and in what follows, bar indicates that the quantity stands for the background,
the indices are raised and lowered by the background metric $\bar g$,
the covariant derivative $\nabla$ is constructed with the background metric,
and the contraction is also understood by that.
This is a solution of the system~\p{action} provided that
\bea
\Lambda_0=\s \Lambda-2(3\a+\b)\Lambda^2.
\label{cos}
\ena
We see that the cosmological constant $\Lambda_0$ should be zero in order to have
a Minkowski spacetime.

\section{Theory around Minkowski spacetime}

In this section, we consider the theory in the flat Minkowski spacetime which is realized for
$\Lambda_0=0$.
We study in what ranges of the parameters the theory~\p{action} becomes unitary and stable
setting $\bar g_{\mu\nu}=\eta_{\mu\nu}$.

For this purpose, we decompose $h_{\mu\nu}$ into their orthogonal parts~\cite{Deser}:
\bea
h_{ij} = 2 \pa_{(i} h_{j)}  + \e_i{}^k \e_j{}^l \phi_{kl}, ~~
h_{0i} = \eta_i + \e_i{}^j \psi_{j}, ~~
h_{00} = n.
\label{dec}
\ena
Subscripts on the indexless variables $(\phi, \eta, \psi)$ denote normalized
spatial derivatives $\pa_i/\small{\sqrt{-\pa_k^2}}$ where $\pa_k^2=\pa_1^2+\pa_2^2$
is the two-dimensional Laplacian.
Gauge invariance of the action~\p{action} allows us to set the three gauge parts
$h_i$ and $\eta$ of the metric to zero by imposing the usual gauge choice
$h_{ij,j}=h_{0i,i}=0$. There remain only the three gauge-invariant components
$(\phi, \psi, n)$ in \p{dec}.

The Einstein tensor has the components
\bea
&& G_{00} = \frac12 \pa_k^2 \phi, \qquad
G_{0i} = \frac12 (\pa_i \dot\phi-\e_i{}^j \pa_k^2 \psi_j), \nn
&& G_{ij} = - \frac12[ \ddot \phi_{ij} - \e_i{}^k\e_j{}^l \pa_m^2 n_{kl}
+ \e_i{}^k \sqrt{-\pa_m^2} \dot \psi_{jk}
+ \e_j{}^k \sqrt{-\pa_m^2} \dot \psi_{ik}],
\ena
where subscripts on $n$ also represent normalized spatial derivatives.
Substituting these into the action~\p{action} and keeping terms up to second order,
we find
\bea
S &=& \int d^3 x \Big[ \frac{\b}{2} \tilde \psi \Box \tilde\psi
+\frac{\s}{2}\tilde\psi^2 + \Big( \a+\frac{\b}{2}\Big) [(\pa_k^2 n)^2 + (\Box\phi)^2]
+ 2 \Big( \a +\frac{\b}{4}\Big) (\pa_k^2 n) (\Box\phi) \nn
&& -\frac{\s}{2} \phi\, \pa_k^2\, n +\frac{1}{2\mu}\tilde\psi(\pa_k^2 n - \Box \phi) \Big],
\label{min}
\ena
where we have defined $\tilde\psi \equiv \pa_i \psi_i =-\sqrt{-\pa_k^2}\psi$
and $\Box \equiv \nabla_\mu^2$.
Because $-\pa_k^2$ is a positive operator, this causes no problem.
There are intrinsically distinct two cases to be discussed separately,
depending on whether $\a + \frac{\b}{2}$ is zero or not.

\subsection{$\a + \frac{\b}{2}\neq 0$ case}

If $\a + \frac{\b}{2}$ is not zero, the Lagrangian from the second order action~\p{min}
can be transformed into
\bea
{\cal L}_2 = \Big(\a + \frac{\b}{2}\Big)\Bigg[ \pa_k^2 n + \frac{(\a + \frac{\b}{4})\phi
-\frac{\s}{4}\phi+\frac{1}{4\mu}\tilde\psi}{\a+\frac{\b}{2}}\Bigg]^2
- \frac{(\a+\frac{3}{8}\b)\b}{\a+\frac{\b}{2}} (\Box\phi)^2
+ \cdots.
\label{min1}
\ena
The first term simply indicates that the non-dynamical field $n$ is determined
in terms of other fields.
The second term tells us that the theory has a dipole
ghost unless $(\a+\frac{3}{8}\b)\b=0$, which gives the first constraint.
So we have to further divide the cases.

\subsubsection{subcase $\a+\frac{3}{8}\b=0$}

In this case, we have $\b \neq 0$ from $\a + \frac{\b}{2}\neq 0$.
Dropping the first term in \p{min1}, our action~\p{min} then gives
\bea
{\cal L}_2 = \tilde \psi \Big( \frac{\b}{2} \Box + \frac{\s}{2} -\frac{1}{2\b\mu^2}\Big)
\tilde\psi
+ \frac{\s}{\b \mu} \tilde \psi \phi - \phi\Big(\frac{\s}{2}\Box +\frac{\s^2}{2\b}\Big)\phi.
\ena
Here we see that inclusion of the LCS term only modifies the mass spectrum,
but does not affect whether the theory contains ghost or not.
In order for the $\tilde\psi$ and $\phi$ fields not to be ghost, we have to have
\bea
\b>0,~~
\s\leq 0.
\label{minc}
\ena

The case $\s=0$ is found by Deser~\cite{Deser}, but the LCS term was not considered there.
In this case, we have only one dynamical mode with mass squared
\bea
m^2 = \frac{1}{(\b\mu)^2}.
\ena

For $\s=-1$, we have two modes with the spectrum
\bea
m_\pm^2 = -\frac{\s}{\b}+ \frac{1}{2(\b\mu)^2} [1 \pm \sqrt{1-4 \s\b\mu^2}],
\ena
which are always positive for \p{minc}. So the theory is also free from tachyons.
The action becomes
\bea
S=\frac{1}{2\kappa_3^2}\int d^3x \left\{ \sqrt{- g} \Big[ \s R
+\b \Big( R_{\mu\nu}^2 -\frac{3}{8} R^2 \Big) \Big] + {\cal L}_{LCS} \right\},
\ena
with the condition~\p{minc}.
This is the new massive gravity~\cite{BHT1} with LCS term.
In the limit $\b \to 0$, one of the spectrum diverges and decouples and we are left
with a single mode with mass
\bea
m^2 = \mu^2.
\ena
This is the well-known topological gravity~\cite{Top}.

\subsubsection{subcase $\b=0$}

Together with $\a+\frac{\b}{2}\neq 0$, we have $\a \neq 0$.
Dropping the non-dynamical first term in \p{min1}, our action~\p{min} then gives
\bea
{\cal L}_2 &=& \Big( \frac{\s}{2} -\frac{1}{16\a\mu^2}\Big) \tilde\psi^2
+ \Big(\frac{\s}{8\a\mu} \phi - \frac{1}{\mu}\Box \phi \Big) \tilde \psi
+ \phi\Big(\frac{\s}{2}\Box -\frac{\s^2}{16\a}\Big)\phi \nn
&=& (\tilde\psi, \phi)
\left(
\begin{array}{cc}
\frac{\s}{2}-\frac{1}{16\a \mu^2} & \frac{\s}{16\a\mu}-\frac{1}{2\mu}\Box \\
 \frac{\s}{16\a\mu}-\frac{1}{2\mu}\Box & \frac{\s}{2}\Box- \frac{\s^2}{16\a}
\end{array}
\right)
\left(\begin{array}{c}
{\tilde\psi} \\
\phi
\end{array}
\right)
\label{min2}
\ena
Diagonalizing this kinetic term, we see that the system always
has modes of opposite norm unless we send $\mu$ to infinity.
This can be most easily checked by taking the determinant of the kinetic
term matrix.
In the limit $\mu\to\infty$, however, the mixing of $\tilde \psi$ and $\phi$
is turned off, and we are left with
\bea
{\cal L}_2 = \frac{\s}{2} \tilde\psi^2
+ \phi\Big(\frac{\s}{2}\Box -\frac{\s^2}{16\a}\Big)\phi.
\ena
Thus we must have
\bea
\s =+1,
\ena
in order to be free from ghost ($\s=0$ gives trivial theory), and
\bea
\a>0,
\ena
in order to be free from tachyon.
This is a special case of $f(R)$ gravity known free from ghosts.
Since higher order terms in $R$ do not affect the quadratic fluctuation
in flat Minkowski spacetime, this conclusion is valid if we include higher orders
in $R$. However the result may change if we consider the theory in
nontrivial backgrounds.

\subsection{$\a + \frac{\b}{2}=0$ case}

Our action~\p{min1} reduces to
\bea
{\cal L}_2 = (\tilde\psi, \pa_k^2 n, \phi)
\left(
\begin{array}{ccc}
\frac{\b}{2}\Box+\frac{\s}{2} & \frac{1}{4\mu} & -\frac{1}{4\mu}\Box \\
\frac{1}{4\mu} & 0 & -\frac{1}{4}(\b\Box+\s) \\
-\frac{1}{4\mu}\Box & -\frac{1}{4}(\b\Box+\s) & 0
\end{array}
\right)
\left(\begin{array}{c}
{\tilde\psi} \\
\pa_k^2 n \\
\phi
\end{array}
\right)
\ena
It is clear that in the limit of $\mu\to\infty$, we have 1 mode of norm
determined by the sign of $\b$ and two modes of opposite norm from the
two linear combinations of $\pa_k^2 n$ and $\phi$. Hence the theory
always has at least one ghost. When $\mu$ term sets in, this conclusion
does not change because these terms do not affect the sign of the highest
power of $\Box$. We thus conclude that there is no unitary theory in this case.

To summarize the result of this section, we have unitary theory around the Minkowski vacuum
for the cases listed in Table~\ref{t1}.
This result agrees with those derived in Refs.~\cite{BHT3,GST} in a slightly different gauge.
\begin{table}[h]
\begin{center}
\caption{Unitary theories around Minkowski vacuum}
\label{t1}
\begin{tabular}{|l|l|l|c|}
\hline
$\a, \b$ & $\s$ &  $\mu$ & number of modes\\
\hline\hline
$\a=-\frac{3}{8}\b, \b> 0$ & $\s=-1$ & arbitrary & 2 (1 for $\mu\to\infty$) \\
\hline
$\a=-\frac{3}{8}\b, \b> 0$ & $\s=0$ & arbitrary & 1 \\
\hline
$\a>0, \b=0$ & $\s=+1$ &  $\mu=\infty$ & 1 \\
\hline
\end{tabular}
\end{center}
\end{table}

\section{Theory around maximally symmetric spacetimes}

We now turn to the study of the general theory around maximally symmetric spacetimes.

Expanding the action~\p{action} around the maximally symmetric spacetimes,
and eliminating $\Lambda_0$ in terms of $\Lambda$ via \p{cos}, we find
that the linear term vanishes due to \p{cos}, and the second order terms give
\bea
{\cal L}_2\!\! &=&\!\! \s \Big[R^{(2)}+ R^{(1)}\frac{h}{2}
+\frac{1}{2} \Lambda(h^2-2h_{\mu\nu}^2) \Big]
+\a \Big[ R^{(1)}{}^2 +12 \Lambda R^{(2)}+6 \Lambda R^{(1)} h
+ 6 \Lambda^2(h^2 -2 h_{\mu\nu}^2)\Big] \nn
&& \hs{-3}+\;\b \Big[ R^{(1)}_{\mu\nu}{}^2 +4 \Lambda \bar g^{\mu\nu} R^{(2)}_{\mu\nu}
-8 \Lambda h^{\mu\nu} R^{(1)}_{\mu\nu} + 2 \Lambda \bar g^{\mu\nu} R^{(1)}_{\mu\nu} h
+ 8 \Lambda^2 h_{\mu\nu}^2 -2 \Lambda^2 h^2 \Big]
+ {\cal L}_{LCS,2},
\ena
where $R_{\mu\nu}^{(1,2)}$ and $R^{(1,2)}$ are defined in the appendix
and ${\cal L}_{LCS}$ is the contribution from the LCS term~\p{cs}.
These can be expressed in terms of
\bea
{\cal G}_{\mu\nu}(h) &\equiv& R_{\mu\nu}^{(1)}-\frac12 R^{(1)} \bar g_{\mu\nu}
-2 \Lambda h_{\mu\nu} \nn
&=& -\frac12[\nabla_\mu \nabla_\nu h-\nabla_\mu h_\nu -\nabla_\nu h_\mu+\Box h_{\mu\nu}
+(\nabla_\la h^\la -\Box h) \bar g_{\mu\nu}] +\Lambda h_{\mu\nu}.
\label{calg}
\ena
Though this looks slightly different from the corresponding one in Ref.~\cite{BHT2},
it is actually the same due to a slight difference in the definition of $R^{(1)}$.
It is not difficult to check that ${\cal G}_{\mu\nu}$ is invariant under the linearized
diffeomorphism:
\bea
\d h_{\mu\nu}=\nabla_\mu \e_\nu+ \nabla_\nu \e_\mu.
\ena

Summing up all terms, the final result is
\bea
{\cal L}_2 = - \Big[ 2 \Lambda (3\a+\b)+\frac{\s}{2}\Big] h^{\mu\nu} {\cal G}_{\mu\nu}(h)
+ \b [{\cal G}_{\mu\nu}(h)]^2 + (4\a+\b) [{\cal G}_\mu^\mu(h)]^2
+ {\cal L}_{LCS,2},
\label{fres}
\ena
where
\bea
{\cal L}_{LCS,2} = \frac{1}{4\mu} \e^{\mu\nu\rho} h_{\mu \la} \nabla_\nu [(\Box h_\rho^\la
-2 \Lambda h_\rho^\la) -\nabla^\la h_\rho ],
\ena
which, thanks to the symmetry of $h_{\mu\nu}$, can be rewritten as
\bea
-\frac{1}{2\mu} \e^{\mu\la\rho} h_{\mu \nu} \nabla_\la {\cal G}_\rho^\nu(h).
\ena

We can rewrite the action by introducing an auxiliary field $k_{\mu\nu}$. We find
\bea
{\cal L}_2 = - \Big[ 2 \Lambda (3\a+\b)+\frac{\s}{2}\Big] h^{\mu\nu} {\cal G}_{\mu\nu}(h)
- \b k^{\mu\nu} {\cal G}_{\mu\nu}(h) -\frac{\b}{4}(k_{\mu\nu}^2-x k^2)
-\frac{1}{2\mu} \e^{\mu\la\rho} h_{\mu \nu} \nabla_\la {\cal G}_\rho^\nu(h),
\ena
where $k=k_\mu{}^\mu$ and
\bea
x \equiv \frac{4\a+\b}{4(3\a+\b)}.
\ena
Indeed, eliminating $k_{\mu\nu}$ by its field equation, we recover the result~\p{fres}.
For $\a=-\frac{3}{8m^2}, \b=\frac{1}{m^2}$ and $\mu =\infty$,
this agrees with the result in \cite{BHT2} without the LCS term.

We take the parametrization~\cite{PS}:
\bea
h_{\mu\nu} = h_{\mu\nu}^T+\nabla_\mu \xi_\nu +\nabla_\nu \xi_\mu
 + \nabla_\mu \nabla_\nu \eta -\frac13 \bar g_{\mu\nu}\Box \eta + \frac13 \bar g_{\mu\nu}h,
\label{para3}
\ena
with
\bea
\nabla^\la h_{\la\mu}^T=0, \qquad
\bar g^{\mu\nu} h_{\mu\nu}^T=0, \qquad
\nabla^\la \xi_\la=0.
\ena
First, substituting~\p{para3} into \p{calg}, we get
\bea
{\cal G}_{\mu\nu} = -\frac12 \Big[(\Box - 2\Lambda) h_{\mu\nu}^T
- \frac13 \nabla_\mu \nabla_\nu \Box \eta + \frac13 \Box(\Box+2 \Lambda)\eta \bar g_{\mu\nu}
+ \frac13 \nabla_\mu \nabla_\nu h - \frac13 (\Box+2\Lambda) h \bar g_{\mu\nu} \Big].
\ena
We then find that \p{fres} gives
\bea
{\cal L}_2 &=& \frac14 h_{\mu\nu}^T (\Box -2 \Lambda)\Big[\{\b(\Box-2 \Lambda)
+4 (3\a+\b)\Lambda+\s\}\bar g^{\mu\rho}\bar g^{\nu\s}
+\frac{1}{\mu}\e^{\mu\la\rho}\bar g^{\nu\s}\nabla_\la \Big] h^T_{\rho\s} \nn
&& +\frac{1}{18}\hat\eta [(8\a+3\b)\Box + 4(3\a+\b)\Lambda-\s]\Box \hat\eta \nn
&& -\frac19 h [(8\a+3\b)\Box + 4(3\a+\b)\Lambda-\s] \sqrt{\Box(\Box+3\Lambda)}\; \hat\eta \nn
&& +\frac{1}{18}h [(8\a+3\b)\Box + 4(3\a+\b)-\s](\Box +3\Lambda) h,
\label{totact1}
\ena
where the field redefinition
\bea
\hat\eta \equiv \sqrt{\Box(\Box+3\Lambda)}\; \s, \qquad
\hat\xi_\mu \equiv \sqrt{\Box+2\Lambda}\; \xi_\mu,
\ena
has been made in order to compensate the Jacobian introduced in changing
field variables from $h_{\mu\nu}$ to \p{para3}~\cite{PS}.
($\xi_\mu$ drops out from the gauge-invariant action here, but we shall have this in
the following discussions of gauge-fixed theory.)

To this action, we add the gauge fixing and the corresponding Faddeev-Popov (FP) ghost terms:
\bea
&& S_{gf} = 
\int d^3 x\sqrt{-\bar g} \Big[-\frac{1}{2a}\chi_\mu
 \bar g^{\mu\nu} \chi_\nu \Big], \nn
&& S_{gh} = -\int d^3 x\sqrt{-\bar g} \bar C^\mu(\d_\mu^\nu \Box
 +\frac{1-b}{2}\nabla_\mu \nabla^\nu+ R_\mu{}^\nu )C_\nu,
\ena
where $a$ and $ b$ are constants, and
\bea
\chi_\nu \equiv \nabla_\mu h^\mu{}_\nu-\frac{b+1}{4} \nabla_\nu h.
\ena
We find
\bea
&&{\cal L}_{gf}=-\frac{1}{2a} \Big[\hat \xi^\mu(\Box+2\Lambda) \hat\xi_\mu
 -\frac49 \hat\eta(\Box+3\Lambda)\hat\eta+\frac{3b-1}{9} h \sqrt{\Box(\Box+3\Lambda)}\hat\eta
-\frac{(3b-1)^2}{144}h\Box h \Big], \nn
&&{\cal L}_{gh}=- \bar V^\mu (\Box+2\Lambda) V_\mu
 + \hat{\bar S}\Big(\frac{3-b}{2}\Box +4\Lambda\Big)\hat S,
\ena
where we have defined~\cite{PS}
\bea
C_\mu \equiv V_\mu +\nabla_\mu S, \quad
\hat S = \sqrt{\Box}\; S, \quad
\nabla_\mu V^\mu=0, \nn
\bar C_\mu \equiv \bar V_\mu +\nabla_\mu {\bar S}, \quad
\hat{\bar S} = \sqrt{\Box}\; \bar S, \quad
\nabla_\mu \bar V^\mu=0.
\ena
The total quadratic Lagrangian is
\bea
{\cal L}_2 &=& \frac14 h_{\mu\nu}^T (\Box -2 \Lambda)\Big[\{\b(\Box-2 \Lambda)
+4 (3\a+\b)\Lambda+\s\}\bar g^{\mu\rho}\bar g^{\nu\s}
+\frac{1}{\mu}\e^{\mu\la\rho}\bar g^{\nu\s}\nabla_\la \Big] h^T_{\rho\s} \nn
&& +\frac{1}{18}\hat\eta \Big[(8\a+3\b)\Box^2 + \Big\{4(3\a+\b)\Lambda-\s
+\frac{4}{a}\Big\} \Box + \frac{12}{a}\Lambda \Big] \hat\eta \nn
&& -\frac19 h \Big[(8\a+3\b)\Box + 4(3\a+\b)\Lambda-\s+\frac{3b-1}{2a}\Big]
\sqrt{\Box(\Box+3\Lambda)}\; \hat\eta \nn
&& +\frac{1}{18}h \Big[(8\a+3\b)\Box^2 + \Big\{(36\a+13\b)\Lambda-\s+\frac{(3b-1)^2}{16a}
\Big\} \Box +3\{4(3\a+\b)-\s\}\Lambda \Big] h \nn
&& - \frac{1}{2a} \hat \xi^\mu(\Box+2\Lambda) \hat\xi_\mu
- \bar V^\mu (\Box+2\Lambda) V_\mu + \hat{\bar S}\Big(\frac{3-b}{2}\Box +4\Lambda\Big)\hat S.
\label{totlag}
\ena

Let us first consider the tensor part.
Clearly there are two kinds of modes in our Lagrangian~\p{totlag};
massless and massive. In order to study the no-ghost condition, we should look at
the propagator and check if the residue at each pole is positive or not.
To do this, let us consider the field equation
\bea
\Big[\{\b(\Box-2 \Lambda) +4 (3\a+\b)\Lambda +\s \}\bar g^{\mu\rho}\bar g^{\nu\s}
+\frac{1}{\mu}\e^{\mu\la\rho}\bar g^{\nu\s} \nabla_\la \Big] h^T_{\rho\s}=0.
\ena
Multiplying this with
$\{ \b(\Box -2\Lambda)+4 (3\a+\b)\Lambda +\s\} \bar g_{\a\mu} \bar g_{\b\nu}
-\frac{1}{\mu}\e_\a{}^\la{}_\mu \bar g_{\b\nu} \nabla_\la$, we get
\bea
\beta^2 (\Box -2\Lambda-M_+^2)(\Box -2\Lambda-M_-^2)h^T_{\a\b}=0,
\label{square}
\ena
where
\bea
M_\pm^2 \equiv  - \frac{4 (3\a+\b)\Lambda +\s}{\b}+\frac{1}{2\b^2\mu^2}
\left[1\pm \sqrt{1- 4\b\mu^2 \{ (12\a+5\b)\Lambda +\s\}} \right] .
\ena
The operator $(\Box-2\Lambda)$ corresponds to the Lichnerowicz operator for
the second-rank tensors in curved spacetime and so its eigenvalues give the masses.
The condition that the propagator has real massive poles in addition to the massless pole
is that
\bea
1- 4\b\mu^2 \{ (12\a+5\b)\Lambda +\s\} \geq 0,
\label{cond1}
\ena
which we assume from now on. We shall see that unitarity and stability of the theory
require that $\b\{(12\a+5\b)\Lambda +\s\}<0$, for which this is satisfied, and then there is
a smooth $\mu\to \infty$ limit.

Thus the propagator for $h_{\mu\nu}^T$ which is given by the inverse of the quadratic
term is found to be
\bea
\frac{ [ \b(\Box -2\Lambda)+ 4(3\a+\b)\Lambda +\s ] \bar g^\a_{(\mu} \bar g^\b_{\nu)}
-\frac{1}{\mu}\e^{\a\la}{}_{(\mu} \bar g^\b_{\nu)} \nabla_\la}{\b^2 (\Box-2\Lambda)
(\Box-2\Lambda-M_+^2)(\Box-2\Lambda-M_-^2)},
\ena
(suitable symmetrization in the indices is understood) which can be decomposed
into three terms
\bea
\frac{A_{1,(\mu\nu)}^{\;\ \a\b}}{\b^2(\Box-2\Lambda)}
+\frac{A_{+,(\mu\nu)}^{\;\ \a\b}}{\b^2(\Box-2\Lambda-M_+^2)}
+\frac{A_{-,(\mu\nu)}^{\;\ \a\b}}{\b^2(\Box-2\Lambda-M_-^2)},
\ena
where
\bea
&& A_{\pm,(\mu\nu)}^{\;\ \a\b} = \frac{ \left[\pm 1+\sqrt{1- 4\b\mu^2 \{(12\a+5\b)\Lambda
+\s\}} \right]\b \bar g^\a_{(\mu}\bar g^\b_{\nu)}\mp 2\b^2\mu \e^{\a\la}{}_{(\mu}
 \bar g^\b_{\nu)}\nabla_\la} {2M_\pm^2\sqrt{1- 4\b\mu^2 \{ (12\a+5\b)\Lambda +\s\}}}, \nn
&& A_1=-(A_++A_-).
\ena
The first term is the same as the contribution from the Einstein-Hilbert action
linearized about the vacuum, and therefore it does not propagate physical degrees
of freedom in three dimensions. So we have to look at the massive poles.
{}From the calculation similar to that in deriving Eq.~\p{square}, we can show
that the eigenvalue of the $\e$ term in $A_\pm$ is $\pm\sqrt{\Box-3\Lambda}$.
We can thus evaluate the residues of the poles at $M_\pm^2$ as follows:
\bea
A_\pm \to \frac{\b\left(\pm 1+\sqrt{1- 4\b\mu^2 \{ (12\a+5\b)\Lambda +\s\}}\right)}
{M_\pm^2 \sqrt{1- 4\b\mu^2 \{ (12\a+5\b)\Lambda +\s\}}}.
\ena
The no-ghost condition from the residue at $M_+^2$ gives
\bea
\b M_+^2>0.
\label{cond2}
\ena
Since $(-1+\sqrt{1- 4\b\mu^2 \{ (12\a+5\b)\Lambda +\s\}}\, )$ is positive or negative
depending on whether $\b\{ (12\a+5\b)\Lambda +\s\}$ is negative or not,
the condition from the pole residue at $M_-^2$ gives $\pm \b M_-^2 >0$ according
to $\mp \b\{ (12\a+5\b)\Lambda +\s\}>0$.
On the other hand, the stability condition requires $M_\pm^2\geq 0$, so \p{cond2} tells us
\bea
\b>0,
\label{cond3}
\ena
and then the lower sign is not allowed. Thus we also have to have
\bea
\b M_-^2>0, \qquad
(12\a+5\b)\Lambda +\s<0.
\label{cond4}
\ena
As $\b>0$, the stability condition implies the unitarity of the theory under the second
condition in \p{cond4}. If the latter is not satisfied, the mode with mass $M_-$
becomes ghost.

We now turn to the scalar part.
The easiest way to see the spectrum for these fields is to take the determinant
of the kinetic term matrix and obtain the eigenvalues of the D'Alembertian.
We find that it is given by
\bea
\frac{9}{(72)^2 a} [(3-b)\Box +8\Lambda]^2 [(8\a+3\b) \Box+4(3\a+\b) \Lambda-\s].
\ena
Consequently there are two modes of mass squared $\frac{8\Lambda}{b-3}$ which are
gauge dependent, and these cancel against the FP ghosts $\hat{\bar S}$ and $\hat S$.
The remaining one is gauge invariant.

We still have to check how the propagator of each mode becomes.
For this purpose, we set $b=1/3$ for simplicity. The relevant part of the action then gives
\bea
{\cal L}_S &=& \frac{1}{18} \left[ h- \hat\eta \sqrt{\frac{\Box}{\Box+3\Lambda}}\, \right]
[(8\a+3\b) \Box + 4(3\a+\b) \Lambda -\s]( \Box + 3\Lambda)
\left[ h- \sqrt{\frac{\Box}{\Box+3\Lambda}}\; \hat \eta \right] \nn
&& + \frac{2}{9a} \hat\eta ( \Box + 3\Lambda)\hat\eta .
\ena
The propagator for $\hat h\equiv \frac13 \left[ h- \sqrt{\frac{\Box}{\Box+3\Lambda}}
\;\hat \eta\right]$ is given by
\bea
&& \frac{1}{[(8\a+3\b) \Box + 4(3\a+\b) \Lambda -\s]( \Box + 3\Lambda)} \nn
&& = \frac{-1}{(12\a+5\b)\Lambda+\s} \Big[\frac{1}{\Box + 3\Lambda}
-\frac{8\a+3\b}{(8\a+3\b) \Box + 4(3\a+\b) \Lambda -\s} \Big].
\ena
The first part and $\hat\eta$ represent the modes cancelling against $\hat S$ and
$\hat{\bar S}$, and the second part is the mode we are left with.
The unitarity condition is thus
\bea
(12\a+5\b)\Lambda+\s>0.
\label{scond1}
\ena
The stability condition is
\bea
\frac{\s-4(3\a+\b)\Lambda}{8\a+3\b} \geq 0 && \mbox{ for AdS}, \nn
\frac{\s-4(3\a+\b)\Lambda}{8\a+3\b} \geq \Lambda && \mbox{ for dS}.
\label{scond2}
\ena

We now notice that the unitarity condition for tensor mode~\p{cond4} and that
for scalar mode~\p{scond1} are incompatible.
This means that we can have either tensor mode or scalar mode.
Thus we must have either
\bea
8\a+3\b=0,
\ena
or
\bea
\b=0 \quad
\mbox{ and } \quad
\mu \to \infty.
\label{cond0}
\ena
The first case corresponds to the decoupling of scalar mode
whereas the second case to the decoupling of the tensor mode.
Let us discuss these cases in turn.

\subsection{$8\a+3\b=0$ case}

This condition was taken as a starting point in \cite{BHT2},
but this is only one of the cases where the theory can be unitary.
In this case, as we have seen above, $\hat \eta$ and $\hat h$ cancel against the FP ghosts
$\hat S$ and $\hat{\bar S}$ and we should concentrate on the tensor part of the action.

Let us now examine the stability condition of the theory for AdS and dS separately.

\subsubsection{AdS case}

We first note that our conditions~\p{cond2}, \p{cond3} and \p{cond4}
confirms a conjecture in Ref.~\cite{BHT5}; noting that the second condition
in \p{cond4} is $\b(\b\Lambda+2\s)<0$ and $\b>0$ in this case, the absence of
tachyon implies the absence of ghost under the condition
$\Omega\equiv -\frac{\b\Lambda+2\s}{2\b\Lambda}<0$
for $\Lambda<0$ in the presence of the LCS term. As conjectured there, the ghost appears
for $\Omega>0$ though the AdS irreps may be unitary. Thus the unitarity
of the irreps is not enough to ensure the unitarity of the field theory, and we need
the off-shell analysis like here, not just field equations, to see this.
In the limit $\mu\to \infty$, $\b M_\pm^2=\b\Lambda/2-\s$, and \p{cond2}
and \p{cond4} both give the same condition in agreement with Ref.~\cite{BHT2}.

Let us next consider the stability condition.
Since $M_+^2 > M_-^2$, only
\bea
M_-^2>0,
\ena
has to be satisfied. This leads to
\bea
1+\b\mu^2(\b\Lambda-2\s) > \sqrt{1-2\b\mu^2(\b\Lambda+2\s)}.
\ena
Under the condition that the left hand side is positive,
\bea
\Lambda> \bar \Lambda \equiv \frac{2\b\mu^2 \s-1}{\b^2\mu^2},
\label{cond5}
\ena
we can take the squares of both sides to obtain
\bea
4\b^2\mu^2\Lambda +(\b\Lambda-2\s)^2>0
\ena
This gives the condition either
\bea
\Lambda >\Lambda_+,\qquad
\mbox{or} \qquad
\Lambda <\Lambda_-,
\label{cond6}
\ena
where we have defined
\bea
\Lambda_\pm \equiv -2\frac{1-\b\mu^2\s\mp\sqrt{1-2\b\mu^2\s}}{\b^2\mu^2}
=-\left(\frac{1\mp\sqrt{1-2\b\mu^2\s}}{\b\mu}\right)^2,
\ena
both of which are negative.
We also have to require \p{cond5}. It turns out that $\bar\Lambda \geq \Lambda_-$.
This excludes the second possibility in \p{cond6}.

Now consider the case $\s=+1$. We must have $\Lambda>\mbox{max}(\Lambda_+,\bar\Lambda)$.
The second condition in~\p{cond4} tells us
\bea
\Lambda<-\frac{2}{\b}.
\ena
However it is easy to show that this is imcompatible with $\Lambda>\Lambda_+$.
Thus $\s=+1$ is excluded. Similarly $\s=0$ is not allowed.

We are left only with the possibility $\s=-1$. In this case,
we find that $\Lambda_+>\bar\Lambda$. So finally we arrive at the condition
\bea
\b>0, \qquad
\s=-1, \qquad
0>\Lambda>\Lambda_+,
\ena
together with arbitrary $\mu$.
In the limit of $\mu\to\infty$, this agrees with the condition in Ref.~\cite{BHT2}.
Our results generalize the condition to more general case.

There is one possible subtlety here when one of the masses vanishes and becomes
degenerate with the graviton. This happens at the boundary of the stability condition:
\bea
\Lambda= \Lambda_\pm.
\ena
This case corresponds to what is known as critical limit.
In the limit $\b\to 0$, we have
\bea
\Lambda_+ \to -(\mu\s)^2,
\ena
which is precisely the case discussed in \cite{LSS} for $\s=+1$.
There appear some additional logarithmic modes which are complicated~\cite{GJ,Log},
and it is argued that the theory is unitary~\cite{LSS}.

\subsubsection{dS case}

For dS, we should have
\bea
M_-^2 > \Lambda.
\ena
It follows that
\bea
M_-^2-\Lambda = \frac{1}{4\b^2\mu^2} \left(1-\sqrt{1-2\b\mu^2(\b\Lambda+2\s)}\right)^2,
\ena
is positive definite, so the stability condition is automatically satisfied.
However we have to impose the condition~\p{cond1} and \p{cond4}.
Both are satisfied for
\bea
\Lambda<-\frac{2\s}{\b},
\ena
but then this tells us that $\s$ must be negative to allow for positive $\Lambda$.
To summarize the conditions, we have
\bea
\b>0, \qquad
\s=-1, \qquad
0<\Lambda<\frac{2}{\b}.
\ena
This result is in agreement with \cite{BHT2,BHT4}.

\subsection{$\b=0$ case}

Let us now turn to the second possibility in~\p{cond0}, which was not considered
in Ref.~\cite{BHT2}. Since the tensor mode does not decouple if $\mu$ is finite,
LCS term should be absent here and the analysis is considerbaly simplified.
We now consider the cases of AdS and dS in turn.

\subsubsection{AdS case $(\Lambda<0)$}

Here we have to require only \p{scond1} and \p{scond2}.
First, for $\a>0$, the conditons give
\bea
-\frac{\s}{12\a} < \Lambda \leq \frac{\s}{12\a}.
\ena
This is possible only for $\s=+1$. Then the conditions are
\bea
\a>0, \qquad
-\frac{1}{12\a} < \Lambda <0, \qquad
\s=+1.
\label{scond3}
\ena

For $\a<0$, we have
\bea
\Lambda< -\frac{\s}{12\a}, \qquad
\mbox{and} \qquad
\Lambda \leq \frac{\s}{12\a}.
\label{scond4}
\ena
Here $\s=\pm 1, 0$ may be all allowed. This is a new possibility compared with
the Minkowski case. However, if we take the limit $\Lambda\to 0$,
these conditions contradict each other and this case ceases to exist.

These results are in agreement with the previous result on Minkowski spacetime.

\subsubsection{dS case $(\Lambda>0)$}

It follows from Eqs.~\p{scond1} and \p{scond2} that for $\a>0$
\bea
-\frac{\s}{12\a}<\Lambda \leq \frac{\s}{20\a}.
\ena
This is possible only for $\s=+1$ to allow for positive $\Lambda$.
The other case $\a<0$ turns out to be inconsistent, so this is the only possibility here.

To summarize the results of this section, we have unitary theory around the maximally
symmetric spacetimes for the cases listed in Table~\ref{t2}.
In the limit of $\Lambda\to 0$, all these results are consistent with the results
for the Minkowski spacetime in the previous section.
\begin{table}[h]
\begin{center}
\caption{Unitary theories around maximally symmetric spacetimes}
\label{t2}
\begin{tabular}{|l|l|l|l|}
\hline
$\a, \b$ & $\Lambda$ & $\s$ &  $\mu$ \\
\hline\hline
$\a=-\frac{3}{8}\b, \b> 0$ & negative, $0>\Lambda>\Lambda_+$ & $\s=-1$ &  arbitrary \\
\hline
$\a=-\frac{3}{8}\b, \b> 0$ & positive, $\frac{2}{\b}>\Lambda>0$ & $\s=-1$ &  arbitrary \\
\hline
$\a>0, \b=0$ & negative, $0> \Lambda>-\frac{1}{12\a}$ & $\s=+1$ &  $\mu=\infty$ \\
\hline
$\a<0, \b=0$ & negative, Eq.~\p{scond4} & all &  $\mu=\infty$ \\
\hline
$\a>0, \b=0$ & positive, $\frac{1}{20\a} \geq \Lambda >0$ & $\s=+1$ &  $\mu=\infty$ \\
\hline
\end{tabular}
\end{center}
\end{table}

\section{Criticality in $4D$}

It is straightforward to extend the calculation of the quadratic fluctuation
around maximally symmetric spacetimes to $D$ dimensions. In four dimensions,
the action~\p{action} without LCS term is the most general as fourth-order action,
since the Riemann tensor squared can be transformed into other terms
using the fact that the Gauss-Bonnet combination is a total derivative.
Since most of the analysis of criticality is done at the level of field equations,
it may be of interest to see how this emerges from the Lagrangian approach.
This provides some new view of the critical theory as to the ghost problem.
Our following discussions are mainly for four dimensions, but we present
the formulae valid for arbitrary dimensions as much as possible.

After some calculation keeping dimensions arbitrary, we find the quadratic Lagrangian
is given by
\bea
{\cal L}_2 = - \Big[ \frac{2(D\a+\b)}{D-2} \Lambda +\frac{\s}{2}\Big] h^{\mu\nu}
 {\cal G}_{\mu\nu}(h)
+ \b [{\cal G}_{\mu\nu}(h)]^2 + \frac{4\a+(4-D)\b}{(D-2)^2} [{\cal G}_\mu^\mu(h)]^2,
\label{freg}
\ena
where
\bea
{\cal G}_{\mu\nu}(h)
\!\!\! &=&\!\!\! -\frac12\Big [\nabla_\mu \nabla_\nu h-\nabla_\mu h_\nu -\nabla_\nu h_\mu
+\Box h_{\mu\nu} +(\nabla_\la h^\la -\Box h) \bar g_{\mu\nu}
-\frac{4}{(D-1)(D-2)}\Lambda h_{\mu\nu} \nn
&& -\frac{2(D-3)}{(D-1)(D-2)}\Lambda h \bar g_{\mu\nu} \Big].
\ena
The condition that the maximally symmetric spacetime is a solution now becomes
\bea
\Lambda_0=\s \Lambda +2(D \a+\b)\frac{D-4}{(D-2)^2}\Lambda^2.
\label{cosd}
\ena

The parametrization~\p{para3} in general dimensions is
\bea
h_{\mu\nu} = h_{\mu\nu}^T+\nabla_\mu \xi_\nu +\nabla_\nu \xi_\mu
 + \nabla_\mu \nabla_\nu \eta -\frac1D \bar g_{\mu\nu}\Box \eta + \frac1D \bar g_{\mu\nu}h,
\label{parad}
\ena
with conditions on the fields similar to three dimensions.
We find
\bea
{\cal G}_{\mu\nu} &=& -\frac12 \Big[\Big(\Box - \frac{2}{(D-1)(D-2)}\Lambda\Big) h_{\mu\nu}^T
- \frac{D-2}{D} \nabla_\mu \nabla_\nu \Box \eta
+ \frac{D-2}{D} \Box \Big(\Box+\frac{2}{D-2} \Lambda\Big) \eta \bar g_{\mu\nu} \nn
&& + \frac{D-2}{D} \nabla_\mu \nabla_\nu h
 - \frac{D-2}{D} \Big(\Box+\frac{4}{D-2}\Lambda\Big) h \bar g_{\mu\nu} \Big].
\ena
A straightforward calculation then yields
\bea
{\cal L}_2 &=& \frac14 h_{\mu\nu}^T \Big(\Box -\frac{4}{(D-1)(D-2)} \Lambda\Big)
\Big[ \b\Big(\Box-\frac{4}{(D-1)(D-2)} \Lambda\Big)
+\frac{4}{D-2} \Lambda(D \a+\b)+\s \Big] h^T_{\rho\s} \nn
&& +\frac{(D-1)(D-2)}{4D^2} \left[ \hat\eta \Delta \Box \hat\eta
- 2 h \Delta \sqrt{\Box\Big(\Box+\frac{2D}{(D-1)(D-2)}\Lambda\Big)}\; \hat\eta\right. \nn
&& \hs{35} \left. + h \Delta \Big(\Box +\frac{2D}{(D-1)(D-2)} \Lambda\Big) h \right],
\label{actiond}
\ena
where we have defined
\bea
&& \hat \eta \equiv \sqrt{\Box\Big(\Box+\frac{2D}{(D-1)(D-2)}\Lambda\Big)}\; \eta, \nn
&& \Delta \equiv \frac{4(D-1)\a+D\b}{D-2}\Box - \frac{4(D-4)}{(D-2)^2}(D\a+\b)\Lambda -\s.
\ena

Our next task is to introduce the gauge fixing and the corresponding FP ghost
terms. This procedure shows that we must have
\bea
4(D-1)\a+D\b=0,
\ena
in order to decouple the scalar modes. As it happens, this is valid for arbitrary
dimensions~\cite{DLLP} though we did not include Riemann tensor squared term.

Now consider the propagator of $h_{\mu\nu}^T$ for $D=4$:
\bea
&& \frac{1}{\Big[\Box-\frac{2}{3}\Lambda\Big]\Big[\b\Big(\Box-\frac{2}{3}\Lambda\Big)
+2(4\a+\b)\Lambda+\s\Big]} \nn
&& = \frac{1}{2(4\a+\b)\Lambda+\s} \left[
\frac{1}{\Box-\frac{2}{3}\Lambda}
-\frac{\b}{\b\Big(\Box-\frac{2}{3}\Lambda\Big)+2(4\a+\b)\Lambda+\s}\right].
\ena
We see that in general there are two modes in our quadratic action~\p{actiond},
and clearly the propagators have residues with positive and negative values
whatever the sign of the factor in front of the two propagators.
If one chooses positive sign for the massless mode from the Einstein theory,
the other gives a mode of negative norm. Since the Einstein mode propagates
in four dimensions in contrast to three, there is no way to make all physical
modes have positive norm. This makes a sharp contrast to three dimensions.
However, it may appear that there is a possibility that they may cancel with
each other if the massive mode becomes massless~\cite{LP},
eliminating ghost modes. This is the critical gravity discussed recently.
The condition is given by
\bea
\a=-\frac{\s}{2\Lambda},
\ena
in agreement with the result in \cite{LP} for $\s=+1$.
However this naive argument for the absence of the ghost mode may not be true
in four dimensions. Even though it looks that the contribution may be cancelled,
actually this critical case introduces new degrees of freedom called log modes~\cite{GJ,Log}.
This could be most easily seen if we go back to our action~\p{actiond},
which shows that the field equation simply becomes
\bea
\Big(\Box-\frac{2}{3}\Lambda\Big)^2 h_{\mu\nu}^T=0.
\ena
This equation certainly contains solutions for the usual Einstein gravity, but
also additional modes which are not solution of the Einstein theory.
In fact there is some discussions on the ghost modes~\cite{PR}.
The question is then whether some boundary conditions can kill the ghosts
and make the theory unitary. This subject deserves further study.

One very interesting feature of the above critical theory is that it can be written
as Einstein plus Weyl squared and cosmological terms:
\bea
S=\frac{1}{\kappa^2}\int d^4x  \sqrt{- g} \s \Big[ R - 2 \Lambda
+ \frac{3}{4\Lambda} C_{\mu\nu\la\rho}^2 \Big],
\ena
up to the Gauss-Bonnet combination which is a total derivative
in four dimensions, where $C_{\mu\nu\la\rho}$ is the Weyl tensor defined by
\bea
C_{\mu\nu\la\rho} = R_{\mu\nu\la\s} -(g_{\mu[\la}R_{\s]\nu}-g_{\nu[\la}R_{\s]\mu})
+\frac13 g_{\mu[\la} g_{\s]\nu} R.
\ena
where the brackets stand for antisymmetrization.
To make the theory close to the Einstein, we should take $\s=+1$.
It is known that the theory of similar structure appears in higher dimensional
critical theories if one includes Riemann tensor squared, and imposes the conditions
that (1) absence of scalar modes, (2) all massless modes, and (3) unique vacuum~\cite{DLLP}.
The last condition is necessary because the equation determining $\Lambda$
in terms of the ``bare'' cosmological constant $\Lambda_0$ becomes quadratic
equation which has two solutions in general.

\section{Discussions and conclusions}

In this paper we have studied for what range of parameters the gravity theory with
higher derivative and topological mass terms can be unitary and stable in
three dimensions, allowing all possible values for the coefficients.
By examining the quadratic fluctuations around possible vacua in the Lagrangian,
we get results summarized in Table~\ref{t1} for the Minkowski spacetime
and in Table~\ref{t2} for maximally symmetric spacetimes.
We have found that the unitarity of the AdS irreps is not enough to ensure
the unitarity of the field theory. Making the off-shell analysis, we have been
able to identify the conditions.
We did this analysis for each of the vacuum separately.
In Ref.~\cite{BHT2}, the unitarity condition around the maximally symmetric spacetimes
was studied with the particular relation $8\a+3\b=0$.
However it is more significant to study these conditions for each case separately
without taking account of constraints from other vacua.
Also to the best of our knowledge, there has not been discussions around maximally
symmetric spacetimes including the topological mass term in the Lagrangian approach.
Considering these possibilities, we have identified all possible conditions
for flat Minkowski and maximally symmetric spacetimes, and found that there is certain
allowed parameter region that has not been explored.

It turns out that only the negative value for the sign of the Einstein term is
allowed if the Ricci tensor squared term is present.
We have seen that without the Ricci tensor squared, the tensor modes decouple
and the theory do not contain spin 2 modes, which may be of little interest as gravity theory.
If the condition requires negative sign for the Einstein term in the presence of
Ricci squared term also for higher dimensions, that would mean that
the low-energy approximation would be different from the Einstein theory.
In analogy with the critical three-dimensional gravity,
it has been argued that there is a possibility that the theory may be unitary
with usual sign of the Einstein term with Ricci squared term~\cite{LP}.
We have seen that the possibility could be understood as cancellation of the
contribution of the mode of negative norm with that of positive norm in our approach.
However there arise logarithmic modes which may be problematic.
If this mode could be eliminated by imposing suitable boundary conditions,
it may give a sensible theory in the tree level, but the question remains
if the theory is unitary and renormalizable at the quantum level.

When the critical condition is imposed, the theory becomes Einstein theory
with Weyl squared and cosmological terms with a special coefficient.
This theory allows the solution in the Einstein theory, but also involves
additional solution. As argued recently in the cosmological context~\cite{Mal},
certain boundary conditions may single out the solutions in the Einstein
theory. Thus Einstein theory may emerge from this kind of higher order theory.

The theory with Weyl squared term may be renormalizable in four dimensions,
but it is not so in higher dimensions. In $D(\geq 4)$ dimensions, we may have
to include $R^{[(D+1)/2]}$ terms, where $R$ stands for the curvature tensors
and the square bracket for the Gauss symbol.
It may be interesting to study what combinations may be allowed in higher dimensions
if the critical conditions are imposed, and see if the theory may become renormalizable.
It has been found that the allowed terms combine into Weyl squared when only
the curvature squared terms are considered and the theory is required to be
critical~\cite{DLLP}. Some study has been done in \cite{LPP,OR}, but more systematic
analysis and determination of the general forms are desirable.
Another interesting question is what happens when dilaton, which always
exists in superstring theories, is included.

We hope to return these problems in the future.

\section*{Acknowledgement}

We would like to thank Paul K. Townsend for numerous valuable discussions
and careful reading of the manuscript, S. Deser, A. Ishibashi and B. Tekin
for very useful discussions.
This work was supported in part by the Grant-in-Aid for
Scientific Research Fund of the JSPS (C) No. 20540283, No.
21$\cdot\,$09225 and (A) No. 22244030.

\appendix

\section{Conventions and useful formulae}

Here we summarize our conventions and formulae necessary in the text.
We give these such that they are valid for any dimension $D$.

Our signature of the metric is $(-,+,\dots)$ and the curvature tensors are given as
\bea
R^\a{}_{\b\mu\nu} &=&
\pa_\mu \G^{\a}_{\b\nu} - \pa_\nu \G^{\a}_{\b\mu}
+ \G^{\a}_{\mu\la} \G^{\la}_{\b\nu} - \G^{\a}_{\nu\la} \G^{\la}_{\b\mu}, \nn
R_{\mu\nu} &=& R^\a{}_{\mu\a\nu}.
\ena

For the background, we take
\bea
\bar G_{\mu\nu}=-\Lambda \bar g_{\mu\nu}, \quad
\bar R_{\mu\nu}= \frac{2}{D-2}\Lambda \bar g_{\mu\nu}, \quad
\bar R_{\mu\nu\rho\la}=\frac{2}{(D-1)(D-2)}\Lambda(\bar g_{\mu\rho}\bar g_{\nu\la}
-\bar g_{\mu\la}\bar g_{\nu\rho}).
\ena
Expansion around the background gives
\bea
\G^\a_{\mu\nu}
&=& \bar \G^\a_{\mu\nu} + \kappa \G^{\a (1)}_{\mu\nu} + \kappa^2 \G^{\a(2)}_{\mu\nu},
\ena
where
\bea
\G^{\a (1)}_{\mu\nu} &=& \frac12 (\nabla_\nu h^\a{}_\mu
 +\nabla_\mu h^\a{}_\nu-\nabla^\a h_{\mu\nu}), \\
\G^{\a(2)}_{\mu\nu} &=& -\frac12 h^{\a\b} (\nabla_\nu h_{\mu\b}
+\nabla_\mu h_{\nu\b}-\nabla_\b h_{\mu\nu}).
\ena
Note that
\bea
\sqrt{-g} = \sqrt{-\bar g} \Big[ 1+\frac{\kappa}{2}h
+ \frac{\kappa^2}{8}(h^2-2 h_{\mu\nu}^2) + O(\kappa^2) \Big].
\ena
We find, to the second order,
\bea
R^\mu{}_{\nu\a\b}\!\! &=&\!\! \bar R^\mu{}_{\nu\a\b} + \kappa R^\mu{}_{\nu\a\b}^{(1)}
+\kappa^2 R^\mu{}_{\nu\a\b}^{(2)}, \nn
R^\mu{}_{\nu\a\b}^{(1)}\!\! &=&\!\! \frac12 (\nabla_\a\nabla_\nu h^\mu_\b
- \nabla_\a\nabla^\mu h_{\nu\b} - \nabla_\b \nabla_\nu h^\mu_\a
+ \nabla_\b\nabla^\mu h_{\nu\a}) +\frac12 \bar R^\c{}_{\nu\b\a} h^\mu_\c
+\frac12 \bar R^\mu{}_{\c\a\b} h^\c_\nu,~~~ \\
R^\mu{}_{\nu\a\b}^{(2)}\!\! &=&\!\!
\nabla_\a \G^{\mu(2)}_{\nu\b} - \nabla_\b \G^{\mu(2)}_{\nu\a}
+ \G^{\mu(1)}_{\la\a} \G^{\la(1)}_{\nu\b}
- \G^{\mu(1)}_{\la\b} \G^{\la(1)}_{\nu\a}.\nonumber
\ena
Similarly
\bea
R^{(1)}_{\mu\nu}\!\! &=&\!\! -\frac12 (\nabla_\mu \nabla_\nu h
- \nabla_\mu h_{\nu} - \nabla_\nu h_\mu + \Box h_{\mu\nu})
+ \frac{2}{(D-1)(D-2)}\Lambda(D h_{\mu\nu} -h \bar g_{\mu\nu}), \nn
R^{(2)}_{\mu\nu}\!\! &=&\!\! \frac12 \nabla_\mu(h^{\b\c} \nabla_\nu h_{\b\c})
-\frac12 \nabla_\b \{h^{\b\c}( \nabla_\mu h_{\nu\c}+ \nabla_\nu h_{\mu\c}
- \nabla_\c h_{\mu\nu}) \} \nn
&& \hs{-10}
- \frac14 (\nabla_\mu h^\b_\a+ \nabla_\a h^\b_\mu -\nabla^\b h_{\a\mu})
(\nabla_\b h^\a_\nu + \nabla_\nu h^\a_\b -\nabla^\a h_{\b\nu})
+ \frac14 \nabla_\a h (\nabla_\mu h^\a_\nu + \nabla_\nu h_\mu^\a -\nabla^\a h_{\mu\nu}), \nn
R^{(1)} \!\! &=&\!\! \nabla_\mu h^\mu -\Box h - \frac{2}{(D-2)} \Lambda h, \nn
R^{(2)} \!\! &=&\!\! \frac12 \nabla_\mu(h^{\b\c} \nabla^\mu h_{\b\c})
-\frac12 \nabla_\b \{h^{\b\c}( 2 h_\c - \nabla_\c h) \} \nn
&& \hs{-3}
- \frac14 (\nabla_\mu h^\b_\a+ \nabla_\a h^\b_\mu -\nabla^\b h_{\a\mu})
(\nabla_\b h^{\a\mu} + \nabla^\mu h^\a_\b -\nabla^\a h_\b^\mu)
+ \frac14 \nabla_\a h ( 2 h^\a -\nabla^\a h) \nn
&&  \hs{-3}
+\frac12 h^{\a\b}\nabla_\a \nabla_\b h
-\frac12 h_\a^\mu \nabla_\b ( \nabla^\a h_\mu^\b +\nabla_\mu h^{\a\b}
-\nabla^\b h^\a_\mu) +\frac{2}{(D-2)}\Lambda h_{\a\b}h^{\a\b},
\ena
where $\Box \equiv \nabla_\mu \nabla^\mu$.
When total derivative terms are dropped, $R^{(2)}$ makes the contribution to the
action
\bea
R^{(2)} \simeq
\frac14 ( h_{\mu\nu}\Box h^{\mu\nu} +h \Box h) +\frac{1}{D-1}\Lambda h_{\mu\nu}^2
+\frac1{(D-1)(D-2)} \Lambda h^2 + \frac12 h_\mu^2.
\ena
We use the notation $\simeq$ to denote equality up to total derivatives.
We also have
\bea
\bar g^{\mu\nu} R^{(1)}_{\mu\nu} &=& \nabla_\mu h^\mu - \Box h, \nn
h^{\mu\nu} R^{(1)}_{\mu\nu} &\simeq& -\frac12(h \nabla_\mu h^\mu
+ h_{\mu\nu}\Box h^{\mu\nu})-h_\mu^2
+ \frac{2}{(D-1)(D-2)}\Lambda( D h_{\mu\nu}^2-h^2),\\
\bar g^{\mu\nu} R^{(2)}_{\mu\nu} &=& \frac{1}{2}h^{\mu\nu} \Big[{\cal G}_{\mu\nu}
+\frac{2}{(D-2)}\Lambda h_{\mu\nu}-\frac{1}{(D-2)}\Lambda h \bar g_{\mu\nu} \Big].
\nonumber
\ena
Note that $\bar g^{\mu\nu} R^{(1)}_{\mu\nu} \neq R^{(1)}$,
because the latter has additional contribution from $h^{\mu\nu} \bar R_{\mu\nu}$.

\end{document}